\documentclass[twocolumn,pre,citeautoscript,superscriptaddress]{revtex4}

\usepackage{amsmath}

\usepackage{graphicx}

\newcommand{\dir}{Figs}
\newcommand{\thickness}{\phi}
\newcommand{\thicknessl}{\phi_{\mbox{\tiny{LdG}}}}
\newcommand{\thicknesse}{\phi_{\mbox{\tiny{el}}}}
\newcommand{\ud}{\mbox{d}}

\begin{document}

\title{Membrane-Protein Interactions in a Generic Coarse-Grained Model for Lipid Bilayers}

\author{Beate West}
\affiliation{Fakult\"at f\"ur Physik, Universit\"at Bielefeld,
        Bielefeld, Germany}
\author{Frank L.H. Brown}
\affiliation{Department of Chemistry and Biochemistry and Department
        of Physics, University of California, Santa Barbara, California
        }
\author{Friederike Schmid}
\affiliation{Fakult\"at f\"ur Physik, Universit\"at Bielefeld,
        Bielefeld, Germany}

\begin{abstract}

We study membrane-protein interactions and membrane-mediated
protein-protein interactions by Monte Carlo simulations of a generic
coarse-grained model for lipid bilayers with cylindrical
hydrophobic inclusions. The strength of the hydrophobic force
and the hydrophobic thickness of the proteins are systematically varied. 
The results are compared with analytical predictions of two
popular analytical theories: The Landau-de Gennes theory and
the elastic theory. The elastic theory provides an excellent description 
of the fluctuation spectra of pure membranes and successfully reproduces 
the deformation profiles of membranes around single proteins.
However, its prediction for the potential of mean force between proteins 
is not compatible with the simulation data for large distances.
The simulations show that the lipid-mediated
interactions are governed by five competing factors: 
Direct interactions, lipid-induced depletion interactions, 
lipid bridging, lipid packing, and a smooth long-range
contribution. The mechanisms leading to ``hydrophobic mismatch'' 
interactions are critically analyzed.

\bigskip


\end{abstract}

\maketitle

\section{Introduction}

Membrane proteins are integral components of biomembranes and account for most 
of the biological processes that take place in and at membranes~\citep{Gennis89}.
Their activity often depends on their distribution within the membrane~\citep{RT01}.
The latter is determined by direct interactions, but also to a significant 
extent by indirect interactions, such as those mediated by the lipid bilayer matrix. 
Lipid-protein interactions are believed to play an important role, {\em e.g.}, 
in controlling the aggregation and activity of gramicidin 
channels~\citep{END83} and rhodopsin~\citep{BHS06}.
Therefore, lipid-mediated interactions between membrane proteins or 
more generally membrane inclusions have been studied intensely for many 
decades~\citep{MB93,GIM98,SMB06}. Natural biomembranes are of course
complex multicomponent systems, and membrane heterogeneities contribute
critically to the lipid-protein interactions~\citep{E03}. However,
a number of mechanisms have been identified that generate lipid-mediated 
protein interactions already in pure, one-component lipid bilayers.

\begin{itemize}

\item[(i)] The mechanism which has been pointed out first in the literature 
is the {\em hydrophobic mismatch interaction}~\citep{BEM91,KI98,DLT99}. 
It arises in situations where the hydrophobic thickness of transmembrane
proteins does not match the equilibrium bilayer thickness. The proteins then 
locally compress or expand the membrane, and the associated free energy penalty 
depends on their distribution in the membrane. This may induce protein clustering.
The effect has been verified experimentally with systematic studies of 
gramicidin~\citep{HHW99} and synthetic model peptides~\citep{SBG02,PK03}.
Theoretically, it has been explained using different continuum theories for 
bilayers~\citep{O78,O79,JM04,MB84,FB93,FBS95,KPD95,BKM03,
H86,H95,HHW99b,NGA98,NA00,DPS93,DPS94,AE96,BB06,BB07}. 
However, it does not seem strong enough to fully account for the experimentally 
observed clustering of proteins in membranes~\citep{DRD02,PBD03}.

\item[(ii)] Even in the absence of hydrophobic mismatch, inclusions locally 
disturb the translational and conformational degrees of freedom of 
lipids~\citep{BEM91,BSBTH96,LZR98,LZR00,LZR01,MBS00,NRG03,KPM04}, which leads to 
local {\em packing interactions}~\citep{M00}. They have been analyzed 
theoretically by sophisticated mean-field studies of effective interactions 
between fully repulsive inclusions inserted in membranes of fixed 
thickness~\citep{LZR00,LZR01,MBS00}. These calculations generally
predict attractive interactions at short distances and repulsive 
interactions at larger distances. 

\item[(iii)] A third class of interactions discussed in the literature
are {\em fluctuation-induced interactions}. Proteins locally affect 
the elastic properties of the membranes -- the bending rigidity 
and/or the spontaneous curvature -- thereby changing the fluctuation
spectrum of the membrane. The corresponding entropy change depends on
the distribution of the proteins, which leads to Casimir 
forces~\citep{GBP03,NP95,GGK96,W01}.

\item[(iv)] In addition to these general interaction mechanisms,
a number of more specific effects have been studied. For example,
lipid-mediated interactions are observed between proteins that
locally induce a strong membrane curvature. The sign of these 
interactions is not clear -- whereas elastic theories predict
repulsion, curvature-inducing proteins have in fact been found to aggregate
in coarse-grained simulations~\citep{RIH07}. This indicates again
that elastic theories alone cannot fully account for the effective
interactions in such a system. Complex interaction mechanisms
have also been predicted for membranes in low-temperature ordered, 
{\em e.g.}, tilted states~\citep{F99}.

\end{itemize}

In the present article, we study the interplay of different 
lipid-mediated interaction mechanisms between simple cylindrical 
inclusions in one-component lipid bilayers by computer
simulations of a generic coarse-grained membrane model. Despite
being very simple, our model reproduces the main phases and 
conformationally driven phase transitions of pure lipid layers, 
including structures as complex as asymmetric and symmetric rippled 
states~\citep{LS07}. It seems therefore suited to study generic 
phenomena that depend on lipid conformations. Specifically, we 
focus on the first two interaction mechanisms listed above, the 
hydrophobic mismatch interaction (i), and the packing 
interactions (ii). The diameters of our inclusions are too 
small to generate measurable Casimir forces, they correspond 
roughly to those of simple $\beta$-helices. The inclusions do 
not induce curvature, and the bilayer is in the fluid state, 
hence additional interactions (iv) also do not contribute. 

In the past decades, a number of computer simulation studies have 
focused on protein-lipid interactions, using both atomistic and 
coarse-grained models (see, {\em e.g.}, Refs.~\citep{ENK04,MVB08} 
for recent overviews). Simulation studies of membrane-mediated 
protein-protein interactions are less abundant. 
Sintes and Baumg\"artner~\citep{SB97} have been
the first to study lipid-mediated interactions in a coarse-grained
molecular model. They considered purely repulsive cylinders immersed 
in a bilayer of lipids which are head-grafted to opposing flat surfaces. 
In qualitative agreement with the mean-field theories cited 
above~\citep{LZR00,LZR01,MBS00}, they find an attractive depletion 
interaction at close distances followed by a repulsive well. 
Smeijers {\em et al.}~\citep{SPM06} have studied the aggregation 
of proteins with different shapes in fusing vesicles. 
Very recently, de Meyer, Venturoli and Smit~\citep{MVB08} 
have presented a systematic study of lipid-mediated protein interactions 
for varying hydrophobic mismatch in a coarse-grained membrane model
with soft DPD (dissipative particle) interactions. Based on their data,
they argue that an important factor driving the hydrophobic mismatch 
interaction is ``hydrophilic shielding'', {\em i.e.}, the influence 
of mismatch on the local arrangement of head groups relative to tails. 

The present study is in many respect complementary to the work of de 
Meyer {\em et al.}~\citep{MVB08}.  First, our model is very 
different: On the one hand, our coarse-grained ``lipid'' structure is much 
simpler, on the other hand, we use hard core, Lennard-Jones type interactions, 
similar to those used in atomistic or systematically coarse-grained 
models~\citep{MRY07}. This allows us to study the influence of 
local lipid packing phenomena, which are almost absent in systems with 
soft DPD potentials, and to diagnose new factors that might contribute 
to the hydrophobic mismatch interaction, such as local chain ordering. 
Second, we systematically vary the hydrophobicity of the protein and study 
its influence on the protein-protein interactions. Third, we relate 
our simulation results to two popular analytical theories of 
lipid-induced interactions: The Landau-de Gennes theory~\cite{O78,O79,JM04}
and the elastic theory for coupled monolayers~\cite{DPS93,DPS94,AE96,BB06,BB07}.
The elastic theory turns out to provide an excellent description of the 
peristaltic and bending fluctuations of pure membranes, and of thickness 
deformation profiles around single proteins. This allows us to analyze
the elastic contribution to the protein-membrane interactions - among other things, 
we identify a mechanism how they may be affected by hydrophilic shielding. 
The simulation results for the membrane-mediated protein-protein interactions, 
however, are not compatible with the predictions of the elastic theory, especially
for large protein distances. In this case the simpler Landau-de Gennes theory
seems to perform better. 

Our paper is structured as follows: After introducing the model
and the simulation method and briefly recollecting the main assumptions 
and predictions of the theories in the next section, 
we present and discuss our simulation results in the section Three.
We conclude with a brief summary.

\section{Models and Methods}

\subsection{Simulation Model and Methods}

We employ a simple generic lipid model~\citep{SDL07} that has been shown 
to reproduce the characteristic high-temperature phase transitions of 
monolayers~\citep{CLS99,DS01} and bilayers~\citep{LS07}. 
The lipids are represented by chains of seven beads with one head 
bead of diameter $\sigma_h$ followed by six tail beads of diameter 
$\sigma_t$.  Beads, that are not direct neighbors in a chain 
interact, {\em via} a truncated and lifted Lennard-Jones potential:
\begin{equation}
  \label{eq:lj_bead}
  V_{\text{bead}}(r) = \left \{
  \begin{array}{rl}
    V_\text{LJ}(r/\sigma) - V_\text{LJ}(r_c/\sigma) & \text{  if $r < r_c$} \\
    0 & \text{  otherwise}    
\end{array}
\right.
\end{equation}
with
\begin{equation}
  \label{eq:lj}
  V_\text{LJ}(x) = \epsilon \left( x^{-12} - 2 x^{-6} \right)
\end{equation}
where $\sigma$ is the mean diameter of the two interacting beads,
$\sigma_{ij} = (\sigma_i + \sigma_j)/2$ ($i,j = h$ or $t$).
Head-head and head-tail interactions are purely repulsive ($r_c = \sigma$) 
while tail-tail interactions also have an attractive
contribution ($r_c = 2\sigma$). Within a ``lipid'' chain beads
are connected to each other by FENE (Finitely Extensible 
Nonlinear Elastic) springs with the spring potential
\begin{equation}
  V_{\text{FENE}} (r) = -\frac{1}{2} \epsilon_{_\text{FENE}} (\Delta r_\text{max})^2
  \log \left (1 - \left ( \frac{r - r_0}{\Delta r_\text{max}} \right )^2
  \right ),
  \label{eq:fene}
\end{equation}
where $r_0$ is the equilibrium distance, $\Delta r_\text{max}$ the maximal
deviation, and $\epsilon_{\text{FENE}}$ the FENE spring constant.
In addition, the chains are given bending stiffness by means of a
bond-angle potential
\begin{equation}
  \label{eq:ba}
  V_{BA}(\theta) = \epsilon_{BA} (1 - \cos(\theta)).
\end{equation}
The aqueous environment of the membrane is modeled with ``phantom'' 
solvent beads~\citep{LS05}, which interact with lipids like head 
beads ($\sigma_s = \sigma_h$), but have no interactions with each other. 
Much like an implicit solvent, the phantom solvent is 
structureless and does not impart unwanted correlations 
onto the bilayers, and it is very cheap from a computational point of 
view. At sufficiently low temperatures it forces the lipids to 
spontaneously self-assemble into stable bilayers~\citep{LS05}. 

Specifically our model parameters are~\citep{DS01,SDL07} $\sigma_h = 1.1 \sigma_t$, 
$r_0 = 0.7\sigma_t$, $\Delta r_\text{max} = 0.2 \sigma_t$, 
$\epsilon_{_\text{FENE}} = 100 \epsilon/\sigma_t^2$, and
$\epsilon_{BA} = 4.7 \epsilon$. The simulations were carried out at 
constant pressure $P = 2. \epsilon/\sigma_t^3$ and temperature 
$k_B T = 1.3 \epsilon$, which is well in the fluid phase region 
of the bilayer. (The bilayer undergoes a main transition to a 
tilted gel phase $L_{\beta'}$ {\em via} a ripple phase 
$P_{\beta'}$ at the temperature $k_B T = 1.2 \epsilon$~\citep{LS07}.)
Comparing the monolayer thickness, $t_0 \sim 3 \sigma_t$, and the area per 
lipid, $a_0 \sim 1.36 \sigma_t^2$, with the corresponding numbers for real 
lipid bilayers in the fluid $L_\alpha$ phase, we find that the values 
in our model roughly reproduce those of DPPC (dipalmitoyl phosphatidylcholine) 
bilayers, if we set $\sigma_t \sim 6$\AA~\citep{o_thesis}. By matching the
temperatures of the main transition ($T_m = 42^\circ$C in DPPC) we
can also identify an energy scale: $\epsilon \sim 0.36 \cdot 10^{-20}$J.

The proteins are modeled as cylinders with the radius $R = 1.5 \sigma_t$,
which have fixed orientation along the bilayer normal (the $z$-axis). We impose
the orientation in order to realize the situation 
considered in the elastic theory as closely as possible -- proteins cannot respond to hydrophobic mismatch 
by tilting. In a real membrane, this would correspond to a situation where the 
orientation of the transmembrane domain of a protein is kept fixed by external 
factors, {\em e.g.}, geometrical constraints in the extramembrane domain.
For comparison, selected simulations were also conducted for proteins
which were allowed to tilt (unconstrained orientations). As in other model
studies~\cite{BB06,MVB08}, the effect of orientation fluctuations on
the results was found to be rather small (see section Three).

The interaction between proteins and lipid or solvent beads 
has a purely repulsive contribution, which is described by a radially 
shifted and truncated Lennard-Jones potential
\begin{equation}
  \label{eq:vrep}
  V_\text{rep}(r) = \left \{
  \begin{array}{rl}
    V_\text{LJ}(\frac{r-\sigma_0}{\sigma}) - V_\text{LJ}(1) 
    & \text{   if $r-\sigma_0 < \sigma$} \\
    0 & \text{   otherwise}    
\end{array}
\right.,
\end{equation}
where $r = \sqrt{x^2+y^2}$ denotes the distance of the interacting partners in 
the $(xy)$-plane, the parameter $\sigma$ is given by $\sigma = (\sigma_t+\sigma_i)/2$ 
for interactions with beads of type $i$ ($i=h,t$, and $s$ for head, tail, and
solvent beads, respectively), $\sigma_0 = \sigma_t$, and $V_\text{LJ}$ has been defined 
above (Eq.~(\ref{eq:lj})). The direct protein-protein interactions have 
the same potential (\ref{eq:vrep}) with $\sigma = \sigma_t$ and $\sigma_0 = 2 \sigma_t$. 

In addition, protein cylinders attract tail beads on a ``hydrophobic'' section of 
length $L$. This is described by an additional attractive potential that depends on 
the $z$-distance between the tail bead and the protein center. The total potential reads
\begin{equation}
  \label{eq:vattr_p}
  V_\text{pt}(r,z) = \epsilon_{pt}\ \Big( V_\text{rep}(r) + V_\text{attr}(r) \cdot W_\text{P}(z) \Big)
\end{equation}
with the attractive Lennard-Jones contribution
\begin{equation}
  \label{eq:vatt}
  V_\text{attr}(r) = \left \{
  \begin{array}{rl}
    V_\text{LJ}(1) - V_\text{LJ}(2) 
    & \text{   if $r-\sigma_0 < \sigma$} \\
    V_\text{LJ}(\frac{r-\sigma_0}{\sigma}) - V_\text{LJ}(2) 
    & \text{   if $\sigma < r-\sigma_0 < 2\sigma$} \\
    0 & \text{   otherwise}    
\end{array}
\right.
\end{equation}
and a weight function $W_\text{P}(z)$, which is unity on a stretch of
length $L-2\sigma_t$ and crosses smoothly over to zero over a distance 
$\sigma_t$ at both sides. Specifically, we use
\begin{equation}
  \label{eq:weight}
  W_\text{P}(z) = \left \{
  \begin{array}{rl}
    1
    & \text{   if $|z| \le l$}\\
    \cos^2[3/2 (|z|-l)]
    & \text{   if $l < |z| < l+\pi/3$} \\
    0 & \text{   otherwise}    
\end{array}
\right.
\end{equation}
with $l = L/2 - \sigma_t$.
The parameter $\epsilon_{pt}$ tunes the strength of the 
lipid-protein interaction, {\em i.e.}, the hydrophobicity of the protein.
It was varied between $\epsilon_{pt}=1$ and $\epsilon_{pt}=6$. 
The hydrophobicity $\epsilon_{pt}=1$ was sufficient to trap the
center of the protein inside the membrane: The fluctuations
of its $z$-position relative to the height of the
membrane $h$ were of the order
$\langle (z_\text{protein} - h)^2 \rangle \stackrel{<}{\sim} 0.5 \sigma_t$ 
for all values of $\epsilon_{pt}$. 
The proteins with unconstrained orientation were modeled as spherocylinders of
length $L$ with essentially the same interaction potentials, except that the
$z$-axis is replaced by the protein axis, and the variable $r$ by the closest
distance to the protein.

The system is studied using Monte Carlo simulations at constant
pressure and temperature with periodic boundary conditions in a 
simulation box of variable size and shape: The simulation box is
a parallelepiped spanned by the vectors $(L_x,0,0), (s_{yx} L_x,L_y,0),
(s_{zx}L_x,s_{zy} L_y,L_z)$, and all $L_i$ and $s_j$ are allowed to
fluctuate. This ensures that the membranes have no interfacial tension.
The system sizes ranged from $200$ to $3200$ lipids, and the simulations 
were run up to $8$ million Monte Carlo steps, where one Monte Carlo 
step corresponds to one Monte Carlo move per bead.
Moves that alter the simulation box were attempted every $50$th
Monte Carlo step. To generate the initial configurations, we set up 
a perfectly ordered bilayer in the $(xy)$-plane with straight 
chains pointing in the $z$-direction and simulated it until it was 
equilibrated, {\em i.e.}, all observables of the system fluctuated
about the equilibrium value and none of the observables shows a trend. 
Typical equilibration times where $1$ million Monte Carlo steps ($4$
million Monte Carlo steps in simulations where we looked at
large-scale height fluctuations). Due to this procedure, the bilayers
were oriented in the $(xy)$-plane.

\begin{figure}[tb]
   \begin{center}
      \includegraphics*[width=3in]{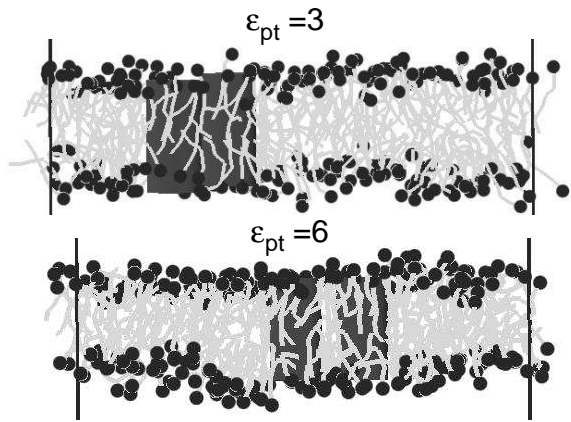}
      \caption{Cross-section snapshot of a model membrane with two inclusions
        of hydrophobic thickness $L=6$ and different hydrophobicity parameter
        $\epsilon_{pt} = 3$ (top) and $\epsilon_{pt} = 6$ (bottom).
        Light grey lines show tail bonds, dark circles the heads (not to scale),
        dark cylinders the inclusions.}
      \label{fig:snapshots}
   \end{center}
\end{figure}

Fig.~\ref{fig:snapshots} shows two snapshots of a system containing two inclusions
whose thickness roughly matches that of the bilayer ($L=6 \sigma_t$), with 
different hydrophobicity parameters $\epsilon_{pt}$. They illustrate the 
existence of membrane-mediated attractive interactions even in the absence of 
hydrophobic mismatch. At low values of $\epsilon_{pt}$, the proteins touch. 
At higher values of $\epsilon_{pt}$, they are separated by a single lipid layer.

Before proceeding to a more quantitative discussion of the simulation results,
we shall now briefly recollect the main assumptions and results of the analytical
theories which we use to analyze our data. 

\subsection{Landau-de Gennes Theory}

One of the oldest theoretical approaches to studying lipid-mediated interactions
between inclusions is based on a Landau-de Gennes expansion of the free energy
in powers of the lipid area or membrane thickness variations~\cite{O78,O79,SM91,JM04}.
In the simplest case, this expansion reads~\cite{JM04}
\begin{equation}
  \label{eq:fe_LdG}
  F_{\mbox{\tiny{LdG}}} = \int \ud^2r \Big\{
  \frac{a}{2} (2\thicknessl)^2 + \frac{c}{2}(2\nabla \thicknessl)^2 \Big\}
\end{equation}
with the boundary condition $\thicknessl = t_R$ at the surface of the inclusion,
where $\thicknessl$ denotes the local deviation of the monolayer thickness
from its equilibrium value $t_0$ in the unperturbed membrane, and $2 (t_R+t_0)$ is the 
hydrophobic thickness of the inclusion. The first term in Eq.~(\ref{eq:fe_LdG}) accounts 
for the area compressibility $k_A$ of the bilayer ($k_A = a (2 t_0)^2$), and the
second term penalizes spatial thickness variations, {\em i.e.}, the variable $c$ is
taken to be positive. Minimizing this free energy for a membrane containing
a single inclusion at $r=0$ yields the deformation profile
\begin{equation}
  \label{eq:profile_LdG}
  \thicknessl(r) = t_R  \frac{K_0(r/\xi)}{K_0(R/\xi)},
\end{equation}
where $R$ is the radius of the inclusion, $\xi = \sqrt{c/a}$ the correlation 
length, and $K_0$ is the modified Bessel function of the second kind. For
$r \xi > 1$, Eq.\ (\ref{eq:profile_LdG}) can be approximated by an
exponential decay
\begin{equation}
  \label{eq:profile_LdG_2}
  \thicknessl(r) \approx t_R  \exp{\left(-\frac{r-R}{\xi}\right)}
 \: \sqrt{\frac{R}{r}}.
\end{equation}
Such exponential laws have often been used to fit data from simulations or molecular 
theories for membranes with inclusions~\cite{SM91,FB93,FBS95,VSS05,V_phd}. 
Eq.~(\ref{eq:fe_LdG}) can also be used to deduce an equation for the monolayer thickness 
fluctuation spectrum
\begin{equation}
\label{eq:tq2_LdG}
\langle |\thicknessl(q)|^2 \rangle =
\frac{k_B T}{4 (a + c q^2)}.
\end{equation}

\subsection{Elastic Theory}

Another popular approach is the elastic theory of coupled monolayers
\cite{H86,H95,HHW99b,HJ90,NGA98,NA00,DPS93,DPS94,AE96,BB06,BB07}. Here the 
membrane is treated as a system of two coupled elastic monolayer sheets.
The basic structure of the different theories is very similar -- they differ
mainly in the choice of the boundary conditions and the number of elastic
terms that they include. For example, early theories have often disregarded 
the possibility that the individual monolayers may have a spontaneous curvature, 
whereas this is usually accounted for in more recent work. Here we shall
use a recent version of the elastic theory developed by Brannigan and 
Brown~\citep{BB06,BB07}, which is fairly ``complete'' in the sense that it
includes all known elastic terms.

We consider a flat lipid bilayer in the $(xy)$-plane. The two constituting 
monolayers are described by four independent fluctuating fields --
two accounting for mesoscopic bending deformations, and two for the 
microscopic protrusions. We assume that the volume of lipids is locally 
conserved, and that the mesoscopic bilayer height and thickness fluctuations
and the protrusions basically decouple.  The latter requirement may seem 
all-too rigid and is actually not imposed in the original model~\citep{BB06},
but it leads to a considerable simplification of the resulting theory
and will be justified {\em a posteriori} by the fact that it describes
the fluctuation spectra of our model bilayers in an excellent way.

The quantities of interest for us are the local bilayer height $h$,
the local monolayer thickness $t$, and the local field $\thicknesse$ which
denotes just the contribution of bending deformations to the thickness,
without the microscopic protrusions. This field is defined in analogy
to the field $\thicknessl$ in Eq.~(\ref{eq:fe_LdG}). With the assumptions 
mentioned above, the spectra of the bilayer height and monolayer thickness fluctuations 
in Fourier space, $\langle |h(q)|^2 \rangle$ and $\langle |t(q)|^2 \rangle$, 
are given by~\citep{BB06}
\begin{align}
\label{eq:hq2}
\langle |h(q)|^2 \rangle =&
\frac{k_B T}{k_c q^4} +
\frac{k_B T}{2 (k_\lambda + \gamma_\lambda q^2)},\\
\label{eq:tq2}
\begin{split}
\langle |t(q)|^2 \rangle =&
\frac{k_B T}{k_c q^4 - 4 k_c \zeta q^2/t_0 + k_A/t_0^2} \\ +&
\frac{k_B T}{2 (k_\lambda + \gamma_\lambda q^2)},
\end{split}
\end{align}
where $k_c$ and $k_A$ are the bending and the compressibility modulus of 
the bilayer, $\zeta$ is related to the spontaneous curvature~\citep{note_zeta},
$t_0$ is the mean monolayer thickness, and the parameters $\gamma_\lambda$
and $k_\lambda$ characterize the protrusions. Fitting our simulation data 
to these theoretical spectra allows us to (i) test the validity of the theory 
and (ii) extract the elastic parameters $k_c$, $k_A$, and $\zeta$, which we 
need for the subsequent analysis of protein-lipid interactions.

Within the decoupling approximation, the protrusions and the height 
fluctuations do not contribute to the protein-induced membrane deformations. 
The free energy of monolayer thickness deformations can be expressed as~\citep{BB07}
\begin{equation}
\label{eq:fdeform}
\begin{split}
F_{el,0} =& \int \ud^2 r \: \bigg\{
\frac{k_A}{2 t_0^2} \thicknesse^2
+ 2 k_c c_0 \nabla^2 \thicknesse
+ 2 k_c \zeta \frac{\thicknesse}{t_0} \nabla^2 \thicknesse
\\
& \qquad + \: \frac{k_c}{2} (\nabla^2 \thicknesse)^2
+ k_G \det (\partial_{ij} \thicknesse)
\bigg\},
\end{split}
\end{equation}
where $(\thicknesse+t_0)$ is the locally smoothed monolayer thickness (without the
protrusions), and we have introduced the spontaneous curvature
of the monolayers $c_0$ and the Gaussian rigidity $k_G$. According to the 
Gauss-Bonnet theorem, the latter only contributes an uninteresting
constant in homogeneous planar sheets and is thus often omitted; in the presence
of inclusions (holes), however, it has to be taken into account~\cite{BB07}.

We consider inclusions with radius $R$ that enforce a certain membrane thickness 
$2t_R$ at their surface. To calculate the deformation profile of the bilayer in the 
vicinity of such an inclusion centered at $r=0$, we minimize the free energy 
$F_{el,0}$ with respect to the profile $\thicknesse(r)$ while keeping the 
membrane deformation at the surface of the inclusion fixed, 
$\thicknesse^\text{surface} \equiv t_R$. This leads to the Euler-Lagrange equation
\begin{equation}
\label{eq:el}
\frac{k_A}{k_c t_0^2} \thicknesse
+ \frac{4 \zeta}{t_0} \nabla^2 \thicknesse
+ \nabla^4 \thicknesse = 0
\end{equation}
with the boundary conditions
\begin{eqnarray}
\label{eq:bc1}
\thicknesse(R) &=& t_R \\
\nabla_r^2 \thicknesse|_R 
&=&
\label{eq:bc2}
- \frac{k_G}{k_c R}t_R'
-2 \left(c_0 + {\zeta}\frac{t_R}{t_0}\right)
\end{eqnarray}
at the surface of the inclusion, and
\begin{equation}
\label{eq:bc3}
\partial_r \thicknesse(r)|_{r \to \infty} = \nabla_r^3 \thicknesse(r)|_{r \to \infty} = 0
\end{equation}
at infinity, where $\nabla^2_r = (1/r) \partial_r r \partial_r$
and $\nabla^3_r = \partial_r \nabla^2_r$, and we have defined
$t_R' = \partial_r \thicknesse|_R$. For a single inclusion,
these equations can be solved analytically, giving~\citep{AE96}
\begin{equation}
\label{eq:profile_solution}
\begin{split}
\thicknesse(r) &=
A_1 J_0 (\alpha_+ r) +
A_2 Y_0 (\alpha_+ r)\\ &+
A_3 J_0 (\alpha_- r) +
A_4 Y_0 (\alpha_- r) 
\end{split}
\end{equation}
with~\citep{note_alpha}
\begin{equation}
\label{eq:alpha}
\alpha_{\pm}
= \sqrt{\frac{2 \zeta}{t_0} \pm 
\sqrt{\left(\frac{2 \zeta}{t_0}\right)^2 - \frac{k_A}{k_c t_0^2}}
},
\end{equation}
where $J_0(x)$ and $Y_0(x)$ are the zeroth order Bessel functions
of the first and second kind, and the coefficients $A_i$ are
determined by the boundary conditions.

The elastic model presented so far only uses material
properties of free, bulk membranes. Inclusions may locally alter the 
lipid properties, {\em e.g.}, the lipid volume, the lipid ordering etc.,
which may in turn affect the elastic  properties ({\em e.g.}, the
equilibrium thickness, the spontaneous curvature, the bending rigidity, 
the compression modulus) of the membrane. In Ref.~\citep{BB07}, 
Brannigan and Brown have demonstrated for the case of varying lipid volume 
that such effects can be incorporated into the theory in a relatively 
straightforward way~\citep{BB07}. 

Here we will discuss this from a more general perspective: Consider some 
scalar quantity $q(r)$ that is distorted from its bulk value $q_0$ by the 
inclusion and locally alters the membrane properties. By symmetry, it will 
introduce two new terms in Eq.~(\ref{eq:fdeform}), 
\begin{eqnarray}
F_{el,q} &=&  F_{el,0} + F_q \qquad \mbox{with}
\nonumber \\
F_q &=& \int \ud^2r \left\{ 
K_1 \frac{\delta q}{q_0} \thicknesse
+ K_2 \frac{\delta q}{q_0} \nabla^2 \thicknesse \right\}.
\label{eq:fq}
\end{eqnarray}
Here $\delta q/q_0$ denotes the relative deviation of $q$, and terms that
do not depend on $\thicknesse$ or that are of higher than quadratic order in the deviations 
$\thicknesse$ and $\delta q/q_0$ have been disregarded.
In general, the additional free energy contribution $F_q$ will change the 
Euler-Lagrange equations, and the solution (\ref{eq:profile_solution}) is no 
longer valid close to the inclusion. The situation however simplifies considerably
if we make the reasonable assumption that the local distortion $\delta q(r)$ 
decays to zero on a length scale which is much smaller than the characteristic 
length scales of the elastic profile. For an inclusion centered at $r=0$, 
we can then replace $\thicknesse(r)$ by $t_R + t_R' (r-R)$ in Eq.~(\ref{eq:fq}), 
and the free energy $F_q$ turns into a surface term,
\begin{equation}
\label{eq:fq_surf}
\begin{split}
{F_q} &= t_R \ K_1 \int_R^\infty 2 \pi \ud r \ r \frac{\delta q(r)}{q_0} \\
&+ t_R'\int_R^\infty \ud r 2 \pi \frac{\delta q(r)}{q_0} (K_1 \: r (r-R) + K_2),
\end{split}
\end{equation}
which only changes the boundary condition (\ref{eq:bc2}): The local distortion 
$\delta q(r)$ renormalizes the spontaneous curvature term in Eq.~(\ref{eq:bc2}) 
according to
\begin{equation}
\label{eq:c0_ren}
\tilde{c_0} = c_0 -
\frac{1}{2 k_c R} \int_R^\infty \ud r \ \frac{\delta q(r)}{q_0} \
(K_1 \ r (r-R)+ K_2).
\end{equation}
Inserting that into Eq.~(\ref{eq:fq}) with (\ref{eq:fdeform}) and exploiting 
the Euler Lagrange equation (\ref{eq:el}), we find that the free energy
of the deformation is given by
\begin{equation}
\label{eq:fdeform_min}
F_{el,q} = \pi k_c R
\Big( t_R \nabla_r^3 \thicknesse|_R - 2 t_R'\big(\tilde{c_0} - \zeta \ t_R/t_0\big) \Big)
+ \mbox{const.},
\end{equation}
where the constant does not depend on the deformation profile.

This result can readily be generalized to situations where several scalar 
quantities $q_\alpha(r)$  affect the membrane simultaneously.
Within our linear approximation, each of them will contribute a separate 
surface term $F_{q_\alpha}$ of the form (\ref{eq:fq_surf}) and the effect on the 
renormalized curvature term (\ref{eq:c0_ren}) will be additive. For given
renormalized curvature $\tilde{c_0}$, Eq.~(\ref{eq:fdeform_min}) still remains valid.

Our findings agree qualitatively with those of Brannigan and Brown in 
Ref.~\citep{BB07}, who also conclude that lipid volume deviations $v(r)/v_0$ at 
the surface of the inclusion effectively renormalize the spontaneous curvature 
term. The actual expression for $\tilde{c_0}$ given in that paper is 
different from ours. The discrepancy is due to an error in the original
analysis.

\subsection{Other Theories}

The elastic theory sketched in the previous section is an
effective interface theory, {\em i.e.}, the main degrees of freedom 
are the positions of the fluctuating interfaces between the monolayers 
and the surrounding solvent. A number of authors have put forward
elastic theories which include the local tilt of chains as supplementary
degrees of freedom in the spirit of the Landau-de Gennes
theory for smectic liquid crystals~\cite{F98,F99,BKM03,FIM06}. 
These theories introduce new elastic parameters which describe,
{\em e.g.}, the splay, twist, and bend modes of the tilt order parameter,
and their coupling to the interfacial degrees of freedom. Due to the
difficulty of accessing the additional parameters, they are not included
in the present analysis.

Another entirely different type of approach is pursued by the molecular 
theories~\cite{FB93,FBS95,HBS97,MBS00,LZR98,LZR00,LZR01}. They
account for the chain character of lipids explicitly and calculate
the packing interactions between proteins and membranes by different
sophisticated mean field methods. Since the results depend on the chain
model, which is different from ours, they can only be compared 
to our simulation data at a qualitative level (see next section).

\section{Simulation Results}

We turn to discuss the simulation results. First, we consider
the properties of pure bilayers, with no inclusions. 
The results confirm the validity of the elastic theory 
for our system, and provide values for the elastic parameters 
that can be used subsequently. In the second step, we
investigate the deformation profiles of a membrane in the vicinity 
of single inclusions. Last, we study the effective 
inclusion-inclusion interactions for inclusions with different 
hydrophobic lengths and hydrophobicity parameters.

\subsection{Elastic Properties of the Pure Membrane}

One of the most powerful approaches to looking at elastic membrane properties 
is the analysis of fluctuation spectra~\citep{GGL99,LE00,MM01,LMK03}.
To determine the spectra for our membranes, we have basically followed the procedure
outlined in Ref.~\citep{LMK03}. We have carried out simulations of a bilayer 
containing $3200$ lipids (and $24615$ solvent beads). The system was divided in 
$N_x \times N_y$ bins in the $(xy)$-plane with $N_x=N_y=20$. In each bin, 
the $z$-value of the mean head position in the $z$-direction was determined for 
both monolayers separately. The average and the difference of the two values 
give the bilayer height spectrum $h(x,y)$ and monolayer thickness spectrum $t(x,y)$, 
respectively.  The two spectra were then Fourier transformed according to 
\begin{equation}
f_{q_x,q_y} = \frac{L_x L_y}{N_x N_y} \sum_{x,y} f(x,y) 
e^{-i (q_x x + q_y y)},
\end{equation}
and the values of $|h_q |^2$ and $|t_q |^2$ were collected in $q$-bins 
of size $0.1$. The binning was made necessary by the fact that the dimensions 
of the simulation box and hence the $q$-values fluctuate. In each bin, the 
averages $\langle |h_q |^2 \rangle$ and $\langle |t_q |^2 \rangle$ were
evaluated.

\begin{figure}[tb]
   \begin{center}
      \includegraphics*[width=3in]{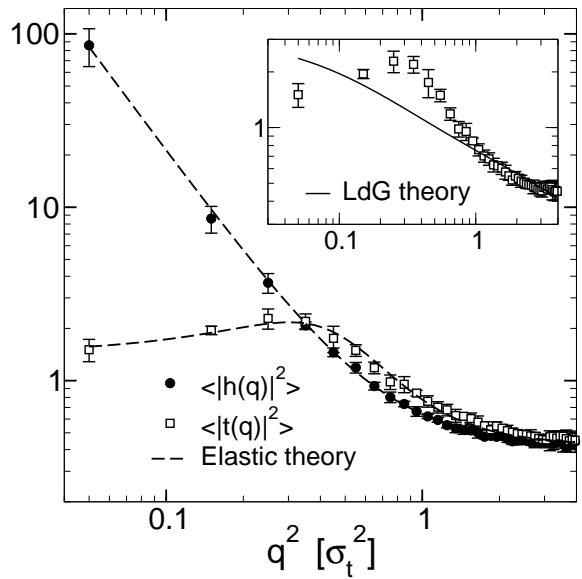}
      \caption{Fourier spectra of height (closed circles) and thickness fluctuations 
        (open squares).  The dashed line shows a fit of the data to the elastic 
        theory~\protect\citep{BB06} (Eqs.~(\protect\ref{eq:hq2}) and (\protect\ref{eq:tq2})).
        The inset shows the thickness data alone with a fit to the 
        Landau-de Gennes theory (Eq.~(\protect\ref{eq:tq2_LdG})).}
      \label{fig:fluctuations}
   \end{center}
\end{figure}

\begin{table}[tb]
  \begin{center}
    \begin{tabular}{|c|c|c|}
      \hline
      Parameter & Value (LJ units) & Value (SI units) \\ \hline
      $k_c$ & $ 6.2 \pm 0.4 \ \epsilon$ & $2.2 \cdot 10^{-20}$ J \\
      $\zeta/t_0$ & $0.15 \pm 0.09 \ \sigma_t^{-2}$ & $0.42$ nm${}^{-2}$ \\
      $k_A/t_0^2$ & $1.3 \pm 0.3 \  \epsilon/\sigma_t^4$ & $3.6 \cdot 10^{-20}$ J/nm${}^4$\\
      $c_0$ & $-0.05 \pm 0.02 \ \sigma_t^{-1}$  & $-0.08$ nm${}^{-1}$ \\
      $k_G$ & $-0.26 \ [-2.8$--$0] \ \epsilon$ & $-1-0 \cdot 10^{-20}$ J\\
      $k_\lambda$ & $1.5 \pm 1 \ \epsilon/\sigma_t^4$ & $4.2 \cdot 10^{-20}$ J/nm${}^4$\\
      $\gamma_\lambda$ & $0.007 \pm 0.01 \ \epsilon/\sigma_t^2$ & $0.7 \cdot 10^{-22}$ J/nm${}^2$
\\
      \hline
    \end{tabular}
    \caption{Elastic constants of our model membrane as obtained from
a fit of the fluctuation spectra of pure membranes to the elastic theory.
Values in SI units are estimates based on the identification
$\sigma_t \sim 6$\AA \ and $\epsilon \sim 0.36 \cdot 10^{-20}$J (see text
for explanation).
}
    \label{tab:elastic}
  \end{center}
\end{table}

The results are shown in Fig.~\ref{fig:fluctuations}. The data illustrates the
characteristic features of the spectra: The height fluctuations are Goldstone modes, 
hence the height spectrum diverges for small wavelength modes. 
In contrast, the monolayer thickness spectrum is limited by the equilibrium 
thickness and tends towards a constant value for small wavevector values~\citep{LE00}.
It exhibits a characteristic peak at $q^2 \sigma_t^2 \sim 0.4$, corresponding
to a soft peristaltic mode with wavelength $\sim 10 \sigma_t$.
At small $q^2$, the fluctuation spectra are dominated by bending deformations.  
For larger values of $q^2$, the spectra are dominated by the protrusion modes and 
are equal for the height and the monolayer thickness fluctuations. 

The solid line in Fig.~\ref{fig:fluctuations} (main frame) shows the fit to the elastic 
theory, Eqs.~(\ref{eq:hq2}) and (\ref{eq:tq2}), with fit parameters $k_c$, $\zeta/t_0$, 
$k_A/t_0^2$, $k_\lambda$,  and $\gamma_\lambda$. The results of the fit are 
given in Table \ref{tab:elastic}. The elastic theory describes the data in an
excellent way.  This confirms the validity of the underlying assumptions, 
most notably, the decoupling approximations (see above). The strength of the coupling 
between bending modes and protrusion modes can be estimated by looking at
the coupling parameter $\gamma_\lambda^2/k_\lambda k_c$. Our fit gives
$\gamma_\lambda^2/k_\lambda k_c = 5 \cdot 10^{-6}$, which is indeed much
smaller than unity. The inset of Fig.~\ref{fig:fluctuations} shows the fit of
the monolayer thickness spectrum to the Landau-de Gennes theory (solid line). 
To make the analyses comparable, we have included a protrusion contribution
and fitted the monolayer thickness deformations to
\begin{equation}
\langle |t(q)|^2 \rangle =
\frac{k_B T}{4 (a + c q^2)} +
\frac{k_B T}{2 (k_\lambda + \gamma_\lambda q^2)},
\end{equation}
(cf. Eq.~(\ref{eq:tq2_LdG})) and the bending deformations to Eq.~(\ref{eq:hq2}) 
simultaneously, with fit parameters $a$, $c$, $k_c$, $k_\lambda$, and $\gamma_\lambda$.
Not suprisingly, the Landau-de Gennes theory cannot reproduce the peak at nonzero $q$ 
in $\langle |t(q)|^2 \rangle$, hence it misses one important characteristic
of the thickness spectrum.

Turning back to the elastic theory,
the analysis of fluctuation spectra yields the values of three 
elastic parameters that are needed in the subsequent analysis:
the bending rigidity $k_c$, the compressibility modulus $k_A$, 
and the extrapolated curvature $\zeta$. The two remaining elastic
parameters in Eq.~(\ref{eq:fdeform}) are the spontaneous curvature 
$c_0$, and the Gaussian rigidity $k_G$. Under the (admittedly bold)
assumption that the two monolayer slabs can be treated as elastic 
continua subject to internal stress, these quantities can be calculated 
from the first and the second moment of the surface tension profile across 
the monolayers~\citep{Safran94} {\em via}
\begin{align}
  \label{eq:curvature}
  k_c \, c_0 &= - \int_0^{\infty} \ud z \ \gamma_\text{int}(z) \ (z-z_0) \\
  \label{eq:gaussian}
  k_G &= 2 \int_0^{\infty} \ud z \ \gamma_\text{int}(z) \ (z-z_0)^2.
\end{align}
Here $\gamma_\text{int}(z)$ denotes the intrinsic surface tension profile, which is 
defined as the difference between the normal and the tangential 
components of the local pressure tensor. The integration starts at the bilayer 
midplane $z=0$, and the reference plane $z=z_0$ is the 
``inextensibility plane'', {\em i.e.}, the plane in which an infinitesimal 
volume element is not compressed or extended if the monolayer is bent. 
For symmetric bilayers with overall tensionless monolayers, one has 
$\int_0^\infty \ud z \ \gamma_\text{int}(z) = 0$, and the result for $c_0$ is
independent of $z_0$. The value obtained for $k_G$, however, depends 
sensitively on the choice of $z_0$.

In the simulation, the pressure tensor is obtained using the virial theorem, 
\begin{equation}
  \mathcal{P}_{\alpha\beta} = 
  \frac{N k_B T}{V} \delta_{\alpha\beta} +
\frac{1}{2V} \left \langle \sum_{i} r^\alpha_{i} F^\beta_{i}\right \rangle,
\label{eq:pressure}
\end{equation}
where $\mathbf{r}_i$ is the position of particle $i$, 
$\mathbf{F}_i$ the force acting on this particle,
$N$ the number of particles, $T$ the temperature, and $V$ the volume.
Because of the periodic boundaries, care must be taken to use a 
version of the expression (\ref{eq:pressure}) that does not depend 
on absolute positions, {\em e.g.}, by rewriting the contribution 
of pairwise forces as $\sum_{i < j} (r_i^\alpha - r_j^\alpha) F_{ij}^{\alpha}$, etc.
The pressure tensor of the whole equilibrated system is diagonal, 
$\mathcal{P}_{\alpha \beta} = \delta_{\alpha \beta} P$, where $P$ is the applied 
pressure. Nevertheless, it varies spatially on the molecular scale. 
In order to measure the local pressure profile, we divide the system
along the $z$-axis in slices of length $\delta z = 0.125 \sigma$. The contributions 
of pairwise forces to the total virial are parcelled out on the bins according
to the Irving-Kirkwood convention~\citep{IK50}, {\em i.e.}, they are distributed
evenly on the line connecting the two interaction partners. The contributions
of our only multibody interactions, the bond-angle potentials (Eq.~(\ref{eq:ba})), 
are distributed evenly on the two participating bonds. Alternatively, 
Goetz and Lipowsky~\citep{GL98} have proposed a method where the contribution of 
multibody forces is spread evenly on all lines connecting the participating partners 
(the bond-angle virials would then be distributed on triangles).
We have also implemented this definition, and found that the effect on the pressure
profile was negligible.
The interfacial tension profile is given by
\begin{equation}
\label{eq:tension}
\gamma(z) = 
\mathcal{P}_{zz}(z)
- \frac{1}{2}(\mathcal{P}_{xx}(z) + \mathcal{P}_{yy}(z)).
\end{equation}

\begin{figure}[tb]
   \begin{center}
     \includegraphics*[width=3in]{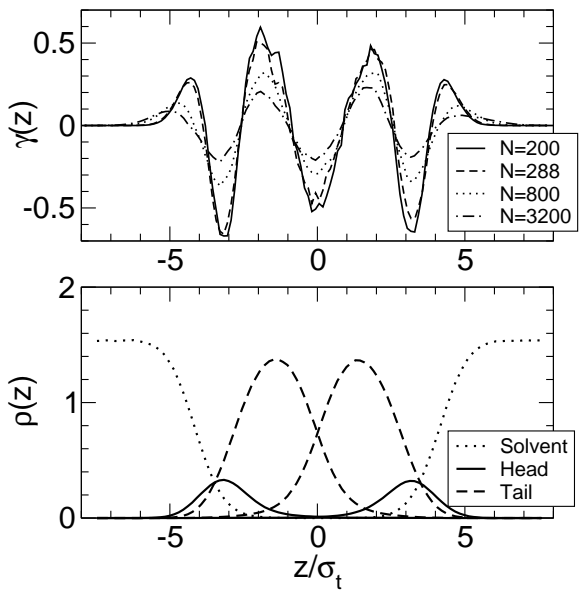}
     \caption{Surface tension profile $\gamma(z)$ for four different system sizes
       with $N=200$, 288, 800, and 3200 lipids (top panel). The bottom panel shows 
       the corresponding density profiles of solvent, head and tail beads
       in the smallest system (200 lipids) for comparison.}
      \label{fig:tension}
   \end{center}
\end{figure}

Figure~\ref{fig:tension} shows the surface tension profiles of a pure
model membrane for four system sizes. Qualitatively, it exhibits the same 
features as the stress profile obtained with the similarly simple 
coarse-grained lipid model of Goetz and Lipowsky~\citep{GL98}. Most 
notably, the surface tension features a negative peak in the bilayer midplane 
(at $z=0$), indicating that the monolayers are strongly bound to each other,
in agreement with atomistic and coarse-grained simulations
of DPPC bilayers~\cite{LE00b,V_phd,MRY07,note_marrink}. 

Due to the height fluctuations, $\gamma(z)$ depends on the lateral system size: 
All profiles are broadened in large systems. In contrast, the expressions 
(\ref{eq:curvature}) and (\ref{eq:gaussian}) are based on a hypothetical 
intrinsic surface tension profile $\gamma_\text{int}(z)$. If such an intrinsic 
profile exists, the actual tension profile  $\gamma(z)$ should be given by 
the convolution of $\gamma_\text{int}(z)$ with the distribution of 
interface heights $W(z')$,
\begin{equation}
\label{eq:convolution}
\gamma(z) = \int \ud z' \ W(z') \ \gamma_\text{int}(z-z').
\end{equation}
In this case, one easily verifies that the integral 
$\Gamma_0 = \int \ud z \: \gamma(z)$ is still zero for tensionless membranes,
and the second moment $\Gamma_2 = \int \ud z \: z^2 \: \gamma(z)$ does not
depend on the shape of the function $W(z)$ for symmetric tensionless bilayers with 
$\int \ud z \:z \:\gamma(z) = 0$. This prediction can be used 
to test the convolution hypothesis, Eq.~(\ref{eq:convolution}). The integral 
$\Gamma_0$ is close to zero in all systems as expected 
($\Gamma_0 \ \sigma_t^2/\epsilon = -0.018, -0.003, -0.04$, and $0.02$ for the system 
with $N=200, 288, 800$, and $3200$ lipids, respectively).
The values of the second moment are $\Gamma_2 / \epsilon = 2.8, 2.7, 2.0$, and $2.8$. 
Hence $\Gamma_2$ does not depend on the system size, which confirms the existence of 
an intrinsic tension profile. 

We note that the integral for the Gaussian rigidity 
(\ref{eq:gaussian}) still depends on the system size, since the lower integration 
bound is finite. Both Eqs.~(\ref{eq:curvature}) and (\ref{eq:gaussian}) are only 
applicable in sufficiently small systems. Therefore, we proceed by evaluating them
in the smallest system with $N=200$ lipids. For the monolayer curvature, we obtain a 
small negative value, $k_c c_0 = -0.3  \pm 0.1 \epsilon/\sigma_t$. This may seem 
surprising, given the fact that the heads in our model are larger than the tail 
beads, and that the tails have to tilt in the low temperature phase (the 
$L_{\beta'}$ phase) in order to accommodate this mismatch. Indeed, the spontaneous 
curvature is found to be positive in the gel state~\citep{note_gel}. At higher 
temperature, the tails disorder and occupy more membrane area, and $c_0$ decreases 
and changes sign as a result. Negative curvatures have also been 
found in more realistic models of DPPC bilayers~\citep{MRY07}. 
The calculated result for the Gaussian rigidity depends on the position 
of the ``inextinsibility plane'' $z_0$, which is not known unambiguously. 
Depending on our choice of $z_0$, we obtain $k_G$ values that range
between $\pm 2.8 \epsilon$. Among these, the positive values can be excluded:
Positive Gaussian rigidity would imply that the bilayers tend to assume 
saddle-shaped configurations, which destabilizes flat bilayer structures on 
large scales and promotes interconnected structures, {\em i.e.}, cubic phases 
or sponge-like structures. Our model bilayers remained flat for all system sizes.
Therefore, only negative values of $k_G$ are physically reasonable. In 
previous work~\citep{BB07}, $z_0$ was taken to be the ``neutral'' plane 
where $\gamma(z)$ crosses zero, {\em i.e.}, $z_0 = 2.6 \sigma_t$ in 
our system (cf. Fig.~(\ref{fig:tension})). Using this value, we obtain
$k_G = -0.26 \epsilon$. 

To summarize this subsection, the Landau-de Gennes theory does not describe
the fluctuations of our model membranes very well. It misses
a soft peristaltic mode in the monolayer thickness spectrum, which is clearly present
in the simulation data. In contrast, the elastic theory fits the data
excellently. We have extracted values for the elastic parameters 
that describe the bending deformations in monolayers from the 
fluctuation spectra and the surface tension profiles of bilayers. 
They are given in Table \ref{tab:elastic}. 
Using the identifications $\sigma_t \sim 6$\AA \ and
$\epsilon \sim 0.36 \cdot 10^{-20}$J (see section 2), we can relate
our model to real lipid bilayers. The values of our elastic parameters
in SI units (see Table \ref{tab:elastic}) have the same order of magnitude 
than those obtained based on all-atom simulations of DPPC~\citep{LE00},
$k_c \sim 4 \cdot 10^{-20}$J,
$k_A/t_0^2 \sim 1.1 \cdot 10^{-20}$J/nm${}^4$,
$\zeta/t_0 \sim 0.18$nm${}^{-2}$~\citep{BB06},
and $c_0 \sim -0.04$ -- $-0.05$ nm${}^{-1}$~\citep{MRY07},
and those based on experimental estimates~\citep{M06},
$k_c \sim 5-20\cdot 10^{-20}$J, 
$k_A/t_0^2 \sim 6 \cdot 10^{-20}$J/nm${}^4$,
$c_0 \sim -0.04$ nm${}^{-1}$, 
and $k_G/k_c \sim -0.8$.

\subsection{Deformation of a Bilayer by a Single Protein}

\begin{figure}[t!]
   \begin{center}
     \includegraphics*[width=3in]{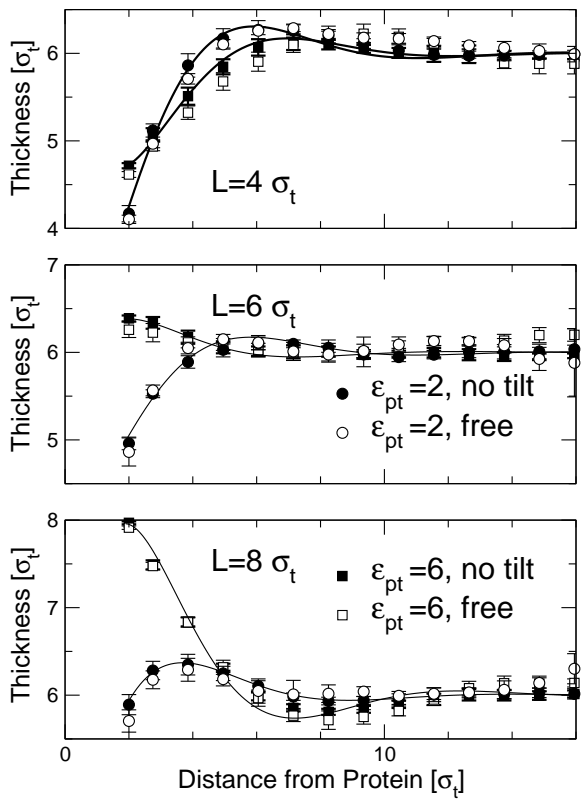}
      \caption{Radial membrane thickness profiles in the vicinity of inclusions with
        different hydrophobic thickness $L$ and hydrophobicity parameter 
        $\epsilon_{pt}$ as indicated. Filled symbols show data for inclusions
        with fixed orientation along the $z$-axis, open symbols correspond
        to unconstrained inclusions. The solid and dashed lines are fits to
        the elastic theory 
        (Eqs.~(\protect\ref{eq:profile_solution}) with (\protect\ref{eq:bc1}), 
        and (\protect\ref{eq:bc2})) with fit parameters $t_R$ and $\tilde{c_0}$.}
      \label{fig:prof_dens}
   \end{center}
\end{figure}

Next we investigate the deformation of a bilayer by a single inclusion.
To this end, we consider the radial profiles of the membrane thickness 
$2 \thickness$, which we define as the mean $z$-distance between the head positions 
in the upper and the lower monolayer. Two parameters are varied, the 
hydrophobicity parameter $\epsilon_{pt}$ (cf. Eq.~(\ref{eq:vattr_p}))
and the hydrophobic thickness $L$ of the inclusion.
The results for $\epsilon_{pt} = 2$ and $6$ are shown in Fig.~\ref{fig:prof_dens} 
for proteins with freely fluctuating orientations and for proteins with orientation 
constrained to the the $z$-axis. The deformation profiles induced by proteins 
with fixed and free orientiations are identical within the error. This is due
to the fact that proteins with free orientations were hardly tilted -- the 
tilt angles were always smaller than $\theta \stackrel{<}{\sim} 0.08$. In the
following, we shall only show data for proteins with fixed orientation.

Looking at Fig.~\ref{fig:prof_dens}, we first observe that the membrane thickness
profiles are not strictly monotonic, but exhibit a characteristic
over- or undershoot at the distance $r \sim 6 \sigma_t$ from the protein.
Such a weakly oscillatory behavior has also been observed in previous
coarse-grained~\cite{VSS05} and atomistic~\cite{CP07} simulations
of protein-induced membrane deformations. In our case, the wavelength
of the oscillation is roughly $\sim 10 \sigma_t$, hence it can be related
to the soft peristaltic mode in the fluctuation spectrum. 

The second observation is that the protein hydrophobicity parameter 
$\epsilon_{pt}$ must exceed a certain value 
in order to produce classical hydrophobic mismatch. If $\epsilon_{pt}$
is too small, the protein effectively repels the lipids, and the membrane
thickness is reduced at the surface of the protein regardless of the value 
of $L$. The hydrophobic section of the protein pins the membrane
thickness for hydrophobicity parameters larger than $\epsilon_{pt} \sim 4$.
This can be rationalized by noting that $\epsilon_{pt} \sim 4$ is
the critical value where touching the protein surface is about as
favorable for tail beads, from an energetic point of view, as being
immersed in the bulk: The maximal contact energy of a tail bead 
in contact with a plane of tail beads is $4 \epsilon$. 

\begin{figure}[t!]
   \begin{center}
     \includegraphics*[width=3in]{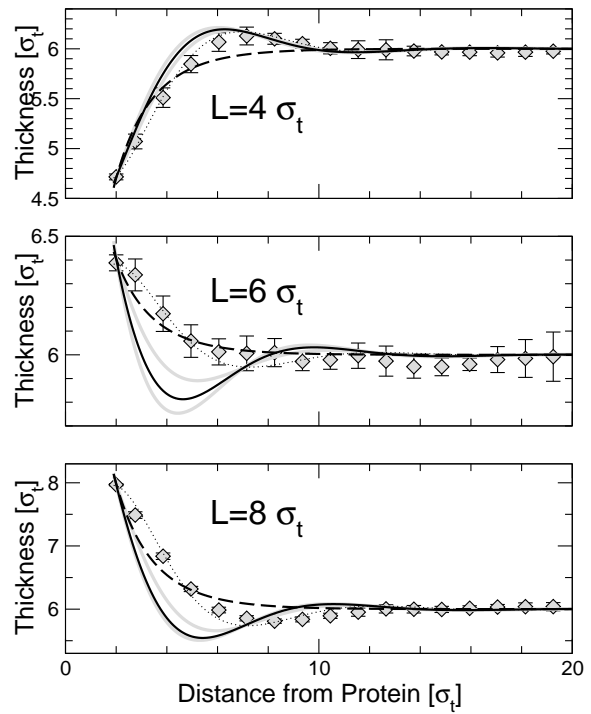}
      \caption{Membrane thickness profiles in the vicinity of an inclusion
        with hydrophobic thickness $L=4 \sigma_t, 6 \sigma_t$, and $8 \sigma_t$, 
        and hydrophobicity parameter $\epsilon_{pt} = 6$, compared
        with fits to the Landau-de Gennes theory (dashed lines), to the
        elastic theory with fixed $c_0 = -0.05\ /\sigma_t$ and 
        $k_G = -0.26 \epsilon$ (solid lines), and to the elastic
        theory with the additional fit parameter $\tilde{c_0}$
        replacing $c_0$ (dotted line). The grey lines indicate the range of
        the fit at fixed $c_0$ if $c_0$ and $k_G$ are varied within
        the error given in Table \ref{tab:elastic}.}
      \label{fig:prof_varfit}
   \end{center}
\end{figure}

We proceed by fitting the profiles to the prediction of the analytical theories.
Since the hydrophobicity of the inclusion must be larger than
$\epsilon_{pt} > 4$ in order to cleave the membrane, we focus on the 
data for $\epsilon_{pt} =6$.  Fig.~\ref{fig:prof_varfit} compares them 
with fits to the prediction of the Landau-de Gennes theory 
(Eq.~(\ref{eq:profile_LdG}) and the elastic theory (Eq.~(\ref{eq:profile_solution}) 
with the boundary conditions Eq.~(\ref{eq:bc2})). In the Landau-de Gennes
case, the three curves were fitted simultaneously with one common fit
parameter $\xi$ and three separate fit parameters $t_R = t_R^{\mbox{\tiny LdG}}$.
Not surprisingly, the Landau-de Genens fit cannot 
reproduce the oscillatory component of the profiles; otherwise, the fit is 
quite reasonable (Fig.~\ref{fig:prof_varfit}, dashed line). 

The thick solid lines in Fig.~\ref{fig:prof_varfit} show the fits to the 
elastic theory with $t_R =: t_R^{\mbox{\tiny el}}$ as sole fit parameter. 
None of them is satisfactory. Hence the ''pure'' version of the elastic theory, 
which explains the profiles in terms of bulk membrane properties only, is 
not sufficient.  In contrast, the data can be fitted very nicely if we release 
the constraint on the value of $c_0$, {\em i.e.}, replace the spontaneous 
curvature by a renormalized curvature $\tilde{c_0}$ (thin dotted lines
in Fig.~\ref{fig:prof_varfit}, solid lines in Fig.~\ref{fig:prof_dens}).
The resulting fit parameter values are essentially the same for 
$\epsilon_{pt} = 6$ (Fig.~\ref{fig:prof_varfit}) and $\epsilon_{pt} = 5$ 
(data not shown) and given in Table \ref{tab:fit_profiles}. 

\begin{table}[tb]
  \begin{center}
    \begin{tabular}{|c|c|c|c|}
      \hline
      $L \ [\sigma_t]$ & 
      $t_R^{\mbox{\tiny LdG}} \ [\sigma_t]$ &
      $t_R^{\mbox{\tiny el}} \ [\sigma_t]$ &
      $\tilde{c_0} \ [\sigma_t^{-1}]$ \\ \hline
      4 & $-0.94 \pm 0.02$ & $-0.66 \pm 0.02$ & $-0.11 \ [-0.12$ -- $-0.10]$ \\
      6 & $ 0.3 \pm 0.02$ & $ 0.18 \pm 0.02$ & $ 0.05 \ [-0.03$ -- $0.06]$ \\
      8 & $ 1.44 \pm 0.02$ & $ 0.93 \pm 0.01$ & $ 0.22 \ [0.15$ -- $0.26]$  \\
      \hline
    \end{tabular}
    \caption{Parameters $t_R^{\mbox{\tiny LdG}}$ and $t_R^{\mbox{el}}$ (monolayer
deformation at the surface of the inclusion) and $\tilde{c_0}$ (renormalized curvature)
obtained from fitting the deformation profiles for large
hydrophobicity parameters $\epsilon_{pt}$ = 5 and 6
to the Landau-de Gennes theory ($t_R^{\mbox{\tiny LdG}}$) and
to the elastic theory ($t_R^{\mbox{\tiny el}}$ and $\tilde{c_0}$).
The remaining fit parameter in the Landau-de Gennes fit is
the decay length $\xi = 2.0 \pm 0.1 \sigma_t$. The uncertainty of 
$\tilde{c_0}$ results from the uncertainty of $k_G$.
}
    \label{tab:fit_profiles}
  \end{center}
\end{table}

From the fit results for $t_R$, we can infer the effective hydrophobic 
thickness $L_\text{eff} = 2(t_0 + t_R)$ of the inclusions. We note that 
the exact relation between $t_R$ and the model parameter $L$ is not
{\em a priori} clear, since the lipid-protein potential is smooth and
varies on the length scale $\sigma_t$. In all cases, the values for 
$L_\text{eff}$ are reasonably close to $L$, {\em i.e.}, well within $1 \sigma_t$.
The fit parameters for $\tilde{c_0}$, however, deviate considerably
from the spontaneous curvature $c_0=-0.05 \pm 0.02 \sigma_t^{-1}$ 
(see Table \ref{tab:elastic}) and depend on $L$. 
This clearly demonstrates that the local structure of the lipids that
surround the inclusion indeed contributes to the boundary conditions, 
as has been discussed in the theory section (Eq.~(\ref{eq:c0_ren})). 

\begin{figure}[t!]
   \begin{center}
     \includegraphics*[width=3.2in]{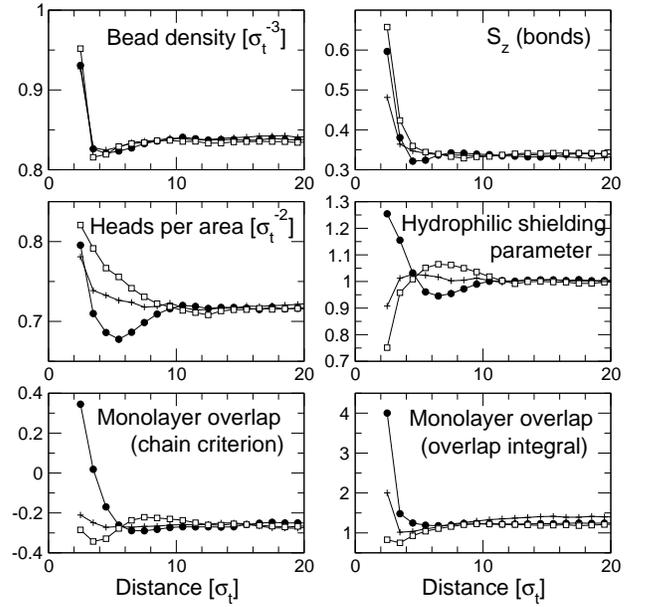}
      \caption{Radial profiles of various quantities as a function of the distance
        from the center of an inclusion with hydrophobic thickness
        $L=4 \sigma_t$ (black circles), $L=6 \sigma_t$ (stars),
        and $L=8 \sigma_t$ (white squares) at $\epsilon_{pt} = 6$.
        Upper left: bead density in the bilayer. Upper right: 
        Nematic order parameter for bonds.
        Middle left: Heads per area in the monolayers.
        Middle right: Hydrophilic shielding parameter (see text
        and Ref.~\protect\citep{MVB08}).
        Lower left and lower right: Monolayer overlap,
        evaluated according to two different prescriptions
        (see text for definitions).}
      \label{fig:prof_other}
   \end{center}
\end{figure}

A similar effect has been observed by Brannigan and Brown in a 
different coarse-grained model in Ref.~\citep{BB07}, and could be
explained satisfactorily by the effect of nonconstant lipid volume. 
The volume per lipid in that model varies substantially over a
range of $r$. In our model, the lipid volume ({\em i.e.}, the
lipid density inside the membrane) is almost constant throughout the system.
Fig.~\ref{fig:prof_other} (upper left) shows profiles of the lipid 
bead density in the membrane for different hydrophobic thicknesses $L$ 
at $\epsilon_{pt} = 6$. Here the lipid bead density $\rho_l$ is defined as 
the number of lipid beads per area divided by the membrane thickness, 
and it is directly related to the local lipid volume $v_l$ {\em via} 
$v_l = n/\rho_l$ with the chain length $n=7$. Apart from an enhancement 
directly at the protein surface, which reflects the attractive 
protein-lipid interaction, and a very shallow depletion zone thereafter, 
the lipid density is nearly constant. Moreover, the curves for different 
hydrophobic thickness $L$ are almost identical. Hence lipid volume 
effects contribute at most an $L$-independent constant to the 
renormalized curvature (\ref{eq:c0_ren}).  

Fig.~\ref{fig:prof_other} also displays profiles of other candidate
quantities that might affect the membrane properties and renormalize the 
curvature term at the surface.  
The upper right panel shows the nematic order parameter for single bonds, 
$S_z = \langle 3 (u_z/u)^2 -1) \rangle/2$, where $u$ refers to the
bond vectors. The profiles show that the nematic order is enhanced in the
vicinity of the inclusion. This is partly related to the increase of
lipid density at the inclusion, since both quantities are correlated.
The details of the nematic order profile, however, cannot be explained
by the density profile alone. The ordering effect is highest for positively
mismatched inclusions. Suprisingly, negatively mismatched inclusions 
induce higher order at the boundary than hydrophobically matching 
inclusions. This may seem counter-intuitive at first, but becomes 
plausible in view of the fact that the monolayers also overlap in the 
vicinity of negatively mismatched inclusions. At larger distances
from the inclusion, the bond order decays monotonically for hydrophobically
matching and positively mismatched inclusions, and exhibits a
non-monotonic dip for negatively mismatched inclusions.

The middle panel in Fig.~\ref{fig:prof_other} shows two quantities that 
are related to the shielding of the hydrophobic membrane interior from 
the solvent. Since shielding is achieved by the heads, the areal head
density gives a direct measure of the effectiveness of shielding.
At constant lipid volume, however, the areal head density depends 
directly on the monolayer thickness $t(r)$ and thus does not qualify as 
independent field $\delta q(r)$ in the elastic theory, {\em i.e.}, it 
cannot contribute directly to Eq.~(\ref{eq:c0_ren}). Instead, we must 
consider a renormalized shielding parameter, such as, {\em e.g.}, 
the ``hydrophilic shielding parameter'' introduced by de Meyer {\em et al.} 
in Ref.~\citep{MVB08}. It is defined as the areal head density divided by
the areal tail density, normalized such that it becomes one far from the 
inclusion. In the original version of the elastic theory, this parameter
would be constant. Our data show that the areal head density is enhanced
directly at the surface of the inclusion for all $L$ 
-- an indirect consequence of the attractive protein-tail interaction.
The subsequent behavior depends on the type of mismatch: For positively 
mismatched inclusions, the areal head density goes up, for negatively
mismatched inclusions, it goes down (Fig.~\ref{fig:prof_other}, middle
left panel). The hydrophilic shielding parameter (middle right panel) 
shows the same behavior at intermediate distances from the inclusion: 
Overshielding for positively mismatched inclusions, undershielding for 
negatively mismatched inclusions. Close to the inclusion, the curves turn 
around, in qualitative agreement with the observations of de Meyer
{\em et al.} in Ref.~\citep{MVB08}. The resulting integral
$\int \ud r \: \delta q(r)/q_0$ (the $K_2$ term in Eq.~(\ref{eq:c0_ren}))
is roughly zero for $L=6 \sigma_t$ and $8 \sigma_t$ and positive for 
$L=4\sigma_t$.

The bottom panel in Fig.~\ref{fig:prof_other} shows profiles of the
monolayer overlap, which measures the amount of interdigitation between
monolayers. It has been calculated following two different conventions:
The left graph shows a chain-related overlap parameter originally introduced 
by Kranenburg {\em et al.} in Ref.~\citep{KVS03}: It is defined as 
$O_{\mbox{\tiny chain}} = \langle 2(l_z - t_0)/l_z \rangle$, 
where $l_z$ is the $z$-component of the end to end vector of 
chains and $t_0$ the monolayer thickness. Far from the inclusion, 
this parameter is negative, indicating that the two monolayers are well 
separated.  Close to the inclusion, it becomes positive for negatively 
mismatched inclusions -- the inclusions pull the lipids inwards and enforce 
a certain amount of interdigitation. For hydrophobically matching 
inclusions and for positively mismatched inclusions, it remains negative 
and roughly constant. The right graph shows a monomer-related overlap
parameter which is defined as the overlap integral of the density profiles
of the two monolayers, $O_{\mbox{\tiny bead}} = \int \ud z 
(\rho_{\mbox{\tiny tail}}^{\mbox{\tiny upper}}(z) -
\rho_{\mbox{\tiny tail}}^{\mbox{\tiny lower}}(z))
$, where $\rho_{\mbox{\tiny tail}}$ denotes the density of tail beads.
The curves for $O_{\mbox{\tiny bead}}$ are qualitatively similar
to those for $O_{\mbox{\tiny chain}}$, except that they are shifted
to positive values -- even far from inclusions, the monolayer densities have
some overlap in the center of the bilayer (cf. Fig.~\ref{fig:tension}).

To summarize this section, we find that the thickness deformation profiles
around single inclusions can be fitted reasonably well with the exponential
law predicted by the Landau-de Gennes theory. The fit with the pure version
of the elastic theory is not good, but the elastic fit becomes much better and 
far superior to the Landau-de Gennes fit, if we allow for the possibility of 
curvature renormalization in the vicinity of the inclusion. We have identified a 
number of quantities which could conceivably contribute to this renormalization,
{\em i.e.}, they vary substantially close to the inclusion and they show a 
sizeable $L$-dependence. However, we cannot pinpoint an obvious single culprit. 
According to Table~\ref{tab:fit_profiles}, the renormalized curvature 
$\tilde{c_0}$ exhibits an almost perfect linear dependence on the hydrophobic 
thickness $L$ of the inclusion: The relation 
$\tilde{c_0} - c_0 \approx 0.078/\sigma_t + 0.0825 (L-2 t_0)/\sigma_t^2$
describes the data in Table \ref{tab:fit_profiles} within $4 \%$. 
None of the quantities shown in Fig.~\ref{fig:prof_other}
produces such a linear behavior in an obvious way. We conclude
that we have either still missed the truly relevant quantity, or that
the observed linear relation is accidental and results from an
interplay of various curvature-renormalizing factors.
>From a physical point of view, it seems likely that most or
all of the quantities discussed above will affect the local
membrane properties, and most notably, the spontaneous curvature: 
Monolayer overlap will favor negative spontaneous curvature, since
the area per tail increases. Variations in the hydrophilic shielding 
parameter will change the local pressure profile and hence affect
the local curvature.
Chain order will favor positive spontaneous curvature, 
since the structure in the chain region resembles that in the gel state, 
where the spontaneous curvature is large and positive.

\subsection{Effective Interactions between Two Proteins}

Finally, we turn to studying the membrane-mediated interactions between 
inclusions.  We study the dilute protein limit, {\em i.e.}, we do not
consider effects from multibody interactions between several proteins.
In order to determine the potential of mean force, we have carried out 
simulations of membranes containing two inclusions and determined their 
radial pair distribution function. The effective potential $w(r)$ is 
\begin{equation}
  \label{eq:epp}
  w(r) = -k_B T \ln g(r),
\end{equation}
where $k_B$ is the Boltzmann constant and $T$ the temperature. Since
the function $g(r)$ varies over several orders of magnitude,
we have used umbrella sampling and reweighting to
determine it accurately at all distances $r$. 

\begin{figure}[h!]
   \begin{center}
     \includegraphics*[width=2.9in]{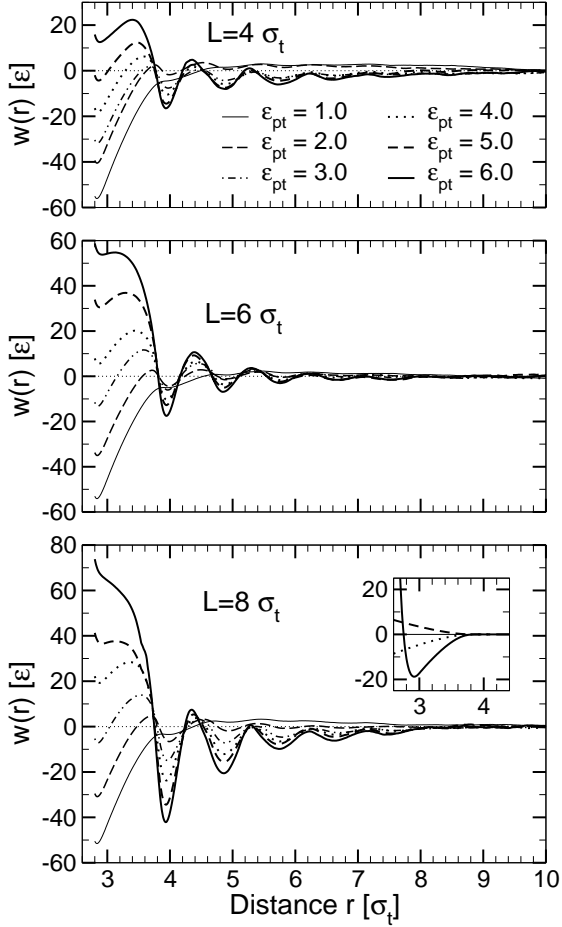}
      \caption{Potential of mean force between two inclusions with hydrophobic 
        thickness $L=4 \sigma_t$ (negative mismatch, top panel),
        $L=6 \sigma_t$ (no mismatch, middle panel), and
        $L=8 \sigma_t$ (positive mismatch, bottom panel) for different
        values of the hydrophobicity parameter
        $\epsilon_{pt}$ as indicated. The inset in the bottom panel
        shows the interactions generated outside of the membrane 
        (solvent-mediated depletion interaction and direct interaction) 
        for hydrophobically matched inclusions (solid line), and the 
        additional contribution of solvent-induced interactions at 
        $L=4 \sigma_t$ (dashed line) and $L=8 \sigma_t$ (dotted line).}
      \label{fig:pmf}
   \end{center}
\end{figure}

The resulting potential curves are shown in Fig.~\ref{fig:pmf}
for several values of the hydrophobicity parameter $\epsilon_{pt}$ 
and the hydrophobic thickness $L$. The most striking feature of
these profiles is their distinctly oscillatory shape. The oscillations have
a wavelength of roughly $1 \sigma_t$ in all systems, which indicates that
they are caused by lipid packing in the vicinity of the inclusions.
Only for the lowest value of the hydrophobicity parameter, $\epsilon_{pt} = 1$, 
the oscillatory structure disappears. Here, we recover qualitatively the 
behavior observed in the simulations of purely repulsive inclusions by Sintes 
and Baumg\"artner~\citep{SB97} and predicted by the corresponding 
molecular mean-field theories~\citep{LZR00,LZR01,MBS00}: The potential 
of mean force exhibits a minimum at close inclusion distances followed 
by a shallow maximum.

Beyond $r > 3.5 \: \sigma_t$, all layering minima obey the general rule 
that they become more shallow with increasing protein separation $r$
and/or decreasing hydrophobicity parameter $\epsilon_{pt}$.
The layering effects are most pronounced at high $\epsilon_{pt}$, indicating 
that lipids pack more tightly if they move closer to the protein surface.
In contrast, the first potential minimum at $r \sim 3 \ \sigma_t$ features
the opposite behavior: It deepens as $\epsilon_{pt}$ decreases and 
disappears for high $\epsilon_{pt}$. 
This minimum results from a number of effects that are not directly 
related to layering: (i) The direct protein-protein interaction 
and the solvent-induced depletion interaction between the hydrophilic 
protein section located outside of the membrane: This contribution is 
roughly constant and can be calculated analytically (inset of 
Fig.~\ref{fig:pmf}).
(ii) A depletion-type interaction induced by the lipids: By pushing the 
proteins towards each other, they maximize their translational and 
conformational entropy. This effect is strongest at low $\epsilon_{pt}$, 
where the protein and the lipids effectively repel each other.
(iii) A bridging interaction induced by the lipids: At higher 
$\epsilon_{pt}$, the lipids gain from being in contact with 
the proteins. Therefore, they tend to squeeze themselves between the proteins,
pushing them apart, and the height of the first minimum goes up.

The competition between the depletion interaction and the lipid bridging effect
accounts for the phenomenon reported earlier (Fig.~\ref{fig:snapshots}), that 
the preferred arrangement of inclusions in the membrane depends on 
$\epsilon_{pt}$: Weakly hydrophobic inclusions tend to touch each other, 
whereas strongly hydrophobic inclusions favor a larger distance where 
they are separated by one single lipid layer.

The influence of hydrophobic mismatch on the potential of mean force can be 
assessed by comparing the potential curves for different hydrophobic thickness $L$.
On the one hand, hydrophobic mismatch affects the local features of 
the potential -- the packing interaction (the strength of the layering)
and the effective contact energy of proteins. On the other hand, it also
contributes a smooth attractive interaction for mismatched proteins with 
$L=4\sigma_t$ and $8\sigma_t$, which superimposes the oscillatory packing
interaction at $r > 4 \sigma_t$. It is worth noting that the sign of 
the additional term is independent of the type of mismatch -- it 
is attractive for both positively and negatively mismatched inclusions.
This smooth long-range contribution to the potential can now be compared
with the predictions of the analytical theories. 

\begin{figure}[t!]
   \begin{center}
     \includegraphics*[width=3in]{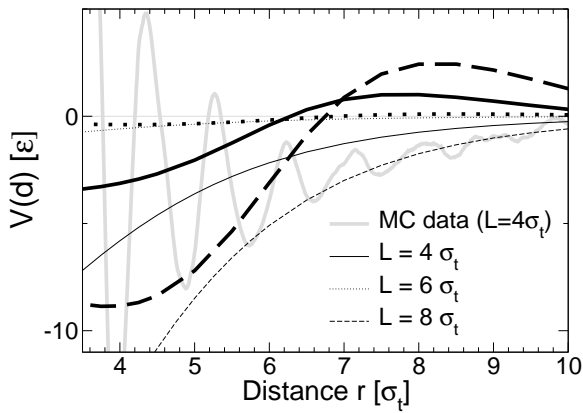}
      \caption{Effective interaction potential between two inclusions according
        to the Landau-de Gennes theory (thin lines) and the elastic
        theory (thick lines) for inclusions with
        different hydrophobic thickness $L$ as indicated. 
        The thick grey line shows the simulation data for $L = 4 \sigma_t,
        \epsilon_{pt}=6$ for comparison.}
      \label{fig:pmf_theory}
   \end{center}
\end{figure}

Fig.~\ref{fig:pmf_theory} shows the corresponding curves for the
Landau-de Gennes theory (thin lines) and the elastic theory (thick lines).
They were calculated numerically by minimizing the free energy --
Eq.~(\ref{eq:fe_LdG}) or Eq.~(\ref{eq:fdeform}) with $\tilde{c_0}$ 
replacing ${c_0}$  -- for systems containing two inclusions at 
given distance $r$, with the boundary condition $\thickness = t_R$
at the surface of the inclusion. The model parameters were taken from
Table \ref{tab:elastic} and \ref{tab:fit_profiles}.
In the Landau-de Gennes calculation, the parameter $4a$ in 
Eq. (\ref{eq:fe_LdG}) was identified with the reduced area 
compressibility coefficient $k_A /t_0^2$ in Table \ref{tab:elastic}.
This may seem inconsistent, since the latter was originally determined from a fit
of the fluctuation spectra to the elastic theory. Unfortunately, the fit
of the spectra to the Landau-de Gennes theory (Fig.~(\ref{fig:fluctuations}))
did not produce dependable parameters $a$ and $c$. The value of $k_A$ in
Table \ref{tab:elastic} is compatible with independent simulation data on
the lipid area increase at finite surface tension~\cite{note_joerg}
and was thus considered to be more reliable. The value for $c$ then follows
from $c = \xi^2 a$, using the value $\xi = 2.0 \sigma_t$ determined from
the fit to the membrane thickness profiles around single inclusions.

To calculate the free energy, we have discretized the corresponding
integrals in real space using a square grid with spatial discretization 
parameter $h$ and a second order difference scheme to evaluate the derivatives. 
The boundary condition was implemented by setting $\thickness = t_R$ inside
the inclusion. (The other boundary condition of the elastic theory, 
Eq.~(\ref{eq:bc2}), follows automatically from the energy minimization). 
The energy was minimized {\em via} a steepest descent method, using
a relaxation scheme introduced in Ref~\citep{FS95}. The final accuracy
was $\int \ud^2 r |\delta F/\delta \thickness| \le 10^{-6}$.
The curves shown in Fig.~\ref{fig:pmf_theory} were obtained using
the spatial discretization $h=0.25 \sigma_t$ and a system of 
size $30 \times 20 \sigma_t^2$ with periodic boundary conditions,
which corresponds to the situation in the Monte Carlo simulations.
Calculations for $h=0.5 \sigma_t$ and system sizes up to 
$70 \times 50 \sigma_t^2$ produced the same curves, hence 
discretization errors and finite size effects are negligible.

Both the Landau-de Gennes theory and the elastic theory predict 
an attractive interaction at distances $r < 6 \sigma_t$, in 
agreement with the trend observed in the simulation data.
For larger distances, the elastic theory predicts the potential
of mean force to become positive and exhibit a peak at
$r \sim 8 \sigma_t$. The simulation data show no indication
for the existence of such a positive peak. Apparently, the elastic
theory fails to reproduce the trend of the simulation 
data at large distances, which is surprising, since this is where
one would expect it to work best. The potential predicted by
the Landau-de Gennes theory is negative and attractive everywhere
and thus better compatible with the data. However, this should
not be overrated, given the overall poorer performance of this
approach when looking at pure membranes and membrane interactions
with single inclusions. The weakly oscillatory behavior of
the potential of mean force in the elastic theory is generated by
the soft peristaltic mode in the fluctuation spectrum, which 
has been shown to leave a clear signature in the shape of 
the distortion profiles around single proteins.
The simulation data suggest that the effect of this mode on
the lipid-mediated interactions between two proteins is destroyed
by some yet unknown mechanism.

\section{Summary and Conclusion}

To summarize, we have determined protein-membrane interactions and 
calculated lipid-mediated interactions between proteins in a simple
generic molecular model for lipid bilayers, and compared the results
to the predictions of two popular continuum theories. Whereas the effect 
of the {\em protein-membrane} interactions on the deformation profiles 
(Fig.~\ref{fig:prof_dens}) can be described very nicely by a
theory that essentially treats the membrane as a pair of
coupled elastic sheets (monolayers), the local lipid structure
clearly dominates the shape of the membrane-induced
{\em protein-protein} interactions.

We note that the specific shape of these packing interactions
depends sensitively on the microscopic details of the system. 
All liquids have a local liquid structure, and packing effects
will clearly also be present in real membranes. However, their
contribution to the potential of mean force will differ
from that in our model. In particular, the amplitude of 
oscillations will presumably be much smaller, since packing effects
are most likely overestimated in our simple Lennard-Jones bead
model. Hence the potentials of mean force shown in Fig.~\ref{fig:pmf}
cannot be related to the potentials of mean force of proteins
in membranes in real systems in a quantitative sense.
At a qualitative level, however, some conclusions can be drawn:

We have identified several factors that may contribute to the
effective potential between cylindrical proteins in our simple 
model membranes: (i) Direct protein-protein interactions, 
(ii) depletion interactions, (iii) lipid bridging interactions, 
(iv) packing interactions, and (v) elastic interactions. 
The interplay and competition of these factors determine the
final, most favorable arrangement of proteins in the bilayer.
Their relative importance is determined by the hydrophobicity
of the proteins and the hydrophobic mismatch. The most dramatic 
effect of hydrophobic mismatch in our system can be associated
with its influence on the factors (ii)-(iv). Thus we do observe 
a ``hydrophobic mismatch interaction'', but the dominant 
contribution to this interaction seems to related to local 
reordering effects, and not to the elasticity of the monolayers.
Intriguingly, the elastic theory does not even describe correctly
the smooth long-range part of the interaction.

As a general trend, the interaction tends to be attractive
{\em i.e.}, it is most favorable for proteins to cluster. 
This is trivially the case for purely oscillatory interactions, but it also 
seems to hold for the additional smooth contribution, regardless 
of the type of mismatch. The observation is consistent with the findings of
de Meyer {\em et al.}~\citep{MVB08} -- the potentials of mean force
presented in that paper are also always attractive, except for
proteins with very large diameters. Whereas protein clustering 
may sometimes be desirable, it also has negative effects - for example, 
it reduces the mobility of the proteins in the membrane, it makes 
them less accessible for other proteins, etc. Our results thus raise 
the question whether the effects reported here have to be 
counteracted by external, not membrane-related protein interactions, 
or whether it is possible to identify mechanisms that 
induce repulsive membrane-mediated interactions between proteins,
and stabilize protein dispersions, {\em e.g.}, in mixed
multicomponent membranes.

\subsection*{Acknowledgment}
We thank Grace Brannigan
and J\"org Neder for 
very helpful discussions. This work was funded by the German Science 
Foundation (DFG) within the SFB 613. F. B. is partially supported by the 
NSF (CHE-0349196). F. B. is a Sloan Research Fellow and a Camille Dreyfus 
Teacher-Scholar. The simulations were mostly carried out at the Paderborn 
center for parallel computing (PC2) and the NIC computer center in J\"ulich.

\vfill

\bibliography{paper}

\begin{thebibliography}{83}
\bibitem[not({\natexlab{a}})]{note_zeta}
The parameter $\zeta$ is given by $\zeta = c_0 - \text{d}c_0/\text{d}\Sigma
  \cdot \Sigma$, where $\Sigma$ is the area per lipid, evaluated for flat
  equilibrium membranes (without inclusions).
\bibitem[not({\natexlab{b}})]{note_alpha}
We note that this expression differs from those given in Refs.\
  \protect\citep{AE96} and \citep{BB07} (which also are mutually different). In
  both papers, the corresponding equations contain typographical errors.
\end{thebibliography}

\end{document}